\documentclass[letterpaper,12pt]{article}

\usepackage{amssymb,latexsym,amsmath}

\usepackage{fullpage}
\usepackage{nicefrac}

\usepackage[pdftex]{graphicx}

\makeatletter \@addtoreset{equation}{section}
\makeatother

\begin{document}
\begin{titlepage}
	\thispagestyle{empty}
	\begin{flushright}
		\hfill{CERN-PH-TH/2009-182}\\
		\hfill{DFPD-09/TH/19}\\
		\hfill{LPTENS 09/30}
	\end{flushright}
	
	\vspace{15pt}
	
	\begin{center}
	    { \LARGE{\bf Universality of the superpotential \\[3mm]
	for  $d=4$ extremal black holes}}
		
		\vspace{20pt}
		
		{Anna Ceresole$^a$, Gianguido Dall'Agata$^{b,c}$, Sergio Ferrara $^{d,e,f}$ and Armen Yeranyan$^{f,g}$}
		
		\vspace{25pt}

		{\small
		{\it ${}^a$ INFN, Sezione di Torino $\&$ Dipartimento di Fisica Teorica\\
		Universit\`a di Torino, Via Pietro Giuria 1, 10125 Torino, Italy}
		
		\vspace{15pt}
		
		{\it ${}^b$ Laboratoire de Physique Theorique de l'Ecole Normale Superieure \\
		24 rue Lhomond, 75231 Paris Cedex 05, France}
		
		\vspace{15pt}
		
		{\it ${}^c$ Dipartimento di Fisica ``Galileo Galilei'' $\&$ INFN, Sezione di Padova \\
		Universit\`a di Padova, Via Marzolo 8, 35131 Padova, Italy}
		
		\vspace{15pt}
		
		{\it ${}^d$ Department of Physics, CERN Theory Division\\
		CH 1211, Geneva 23, Switzerland}
		\vspace{15pt}

		{\it ${}^e$
		Department of Physics and Astronomy \\
		University of California, Los Angeles, CA, USA}
		
		\vspace{15pt}
		
		{\it ${}^f$ INFN - LNF,
		Via Enrico Fermi 40, I-00044 Frascati, Italy}
		
		\vspace{15pt}
		
		{\it ${}^g$ Department of Physics, Yerevan State University, \\
		Alex Manoogyan St. 1, Yerevan, 0025, Armenia}}
		
		\vspace{30pt}
		
		{ABSTRACT}
	\end{center}
	
	\vspace{10pt}
	
We provide a general strategy to obtain the superpotential $W$ for both BPS and non-BPS extremal black holes in $N=2$ four dimensional supergravities based on symmetric spaces.
This extends the construction of $W$ in terms of U-duality invariants that was presented in previous work on the $t^3$ model.
As an application, we explicitly provide $W$ and the solutions to the related gradient flows for the $st^2$ and $stu$ models.
The procedure is shown to hold also for the full $N=8$ theory.
The role of flat directions in moduli space is clarified.

\end{titlepage}

\baselineskip 6 mm

\section{Introduction}

The effective potential for $d=4$, $N=2$ extremal black holes is given in terms of a superpotential $W$ that is known to drive the first order flow equations for the radial evolution of the scalar fields $z^i$ ($i=1,\ldots, n$) and the warp factor $U$ from asymptotic infinity towards the black hole horizon \cite{SUSY,Ceresole:2007wx}:
\begin{eqnarray}
V_{BH} &=& W^2+4g^{i{\bar\jmath}}\partial_i W \partial_{\bar\jmath} W\, ,\\[2mm]
U^\prime &=&-e^UW\, , \label{VBH}\\[2mm]
z^{\prime i}&=&-2e^Ug^{i{\bar\jmath}}\partial_{\bar\jmath} W\, . \label{zprime}
\end{eqnarray}
The real scalar function $W(z^i,\bar z^i)$ also gives the entropy of the  extremal black hole at the horizon $S_{BH}=\pi W^2$,  the $ADM$ mass, $M_{ADM}=W$, and the scalar charges at infinity, $\Sigma_i=\partial_i W$.
For BPS solutions, $W$ coincides with the modulus of the central charge $Z(z,\bar z;q,p)$ of the supersymmetry algebra \cite{SUSY}.
For the non BPS branch, $|Z|$ is replaced by a real function $W(z,\bar z)$ of the scalar fields $z^i$ and of the electric/magnetic charges $(q,p)$ \cite{Ceresole:2007wx}.
Unfortunately this function does not seem to have such a clear cut algebraic meaning and its form has been found mostly on a case by case basis \cite{Ceresole:2007wx,Andrianopoli:2007gt,Cardoso5d,varinonBPS,Bellucci:2008sv}.
However, it was proved in \cite{Andrianopoli:2009je} that, like the black hole potential $V_{BH}$ itself and like the central charge $|Z|$, also the ``fake superpotential'' $W$ should be given by a duality invariant expression.
A rather non-trivial test of this statement was recently given in \cite{Ceresole:2009iy}, where, by analyzing the $t^3$ model, it was proved that $W$ could be a complicated non-polynomial function of the basic duality invariants.
Similar results were also obtained by analyzing the black hole evolution equations obtained by performing a time-like reduction to 3 dimensions in \cite{Bossard:2009we}, where it was shown that, for general non-BPS extremal black holes in $N = 8$ supergravity, $W$ is obtained by solving a polynomial equation of degree 6 in $W^2$ whose coefficients are SU(8) invariant
functions of the central charges.

This paper further extends these results by showing that  $W$ enjoys the universal property of being a function of the duality invariants of the underlying special geometry. Moreover, it shows that U-duality invariance of extended supergravity constrains the form of $W$, and it provides a tool to construct it explicitly in broad classes of models.
In special geometry, duality invariant quantities are those that remain unchanged under the simultaneous action of the duality group on the charge vector $Q = (p^\Lambda, q_\Lambda)$ and on the scalar fields (expressed through the symplectic sections $(X^\Lambda,F_\Lambda)$, with $\Lambda=(0,i)=1,\ldots,n+1$).

It was already known for some time that the geometric properties of  the scalar  manifolds and U-duality allow to classify a priori the solutions to the attractor equations and their supersymmetry properties \cite{orbits,orbits2}.
Among the special geometries based on cubic $F$ functions, those that are based on a symmetric coset space $G/H$, enjoy a full classification \cite{Cremmer:1984hc}.
A good representative of this family is the $stu$ model \cite{triality,Behrndt:1996hu}, which has always been used as a working example to understand generic physical and mathematical features and which has already been thoroughly investigated in the context of the first order formalism in \cite{Bellucci:2008sv}.
This model can be seen as a sector of the maximally extended 4-dimensional $N=8$ supergravity, where, beside the $28$ vectors, there are $70$ scalar fields spanning the coset space E$_{7(7)}/$SU(8), with U-duality symmetry E$_{7(7)}$.
The $stu$ model is the consistent truncation to 4 vectors and 3 complex scalars, each spanning the coset SU(1,1)/U(1).
Interestingly, it can also be viewed simply as an $N=2$ supergravity model coupled to 3 vector multiplets plus the graviphoton, being a fourth vector.

In \cite{Ceresole:2009iy} it was shown that for the $t^3$ model, obtained in a suitable limit from  the $stu$ model, one can express $W$ purely as a combination of U-duality invariants.
Our present aim is to provide a universal procedure to construct $W$ in $N=2$ special geometries and to understand how flat directions enter the game.
To this end, we focus on minimal symmetric  cosets of  rank 1, 2 and 3 (the minimal symmetric coset is SU(1,1)/U(1), underlying the $t^3$, $st^2$ and the $stu$ models).
We shall put in evidence that, while the $t^3$ model has no flat directions, for $st^2$ and $stu$ the space of flat directions has rank 1 and rank 2 respectively.
When one considers larger cosets, as classified in \cite{Ferrara:2007tu}, one sees that the dimensions of the cosets and the number of flat directions are increased, but NOT their rank.
Therefore what we present here is representative of the whole class of symmetric spaces.
Furthermore, our method has a natural interpretation also in connection with the full $N=8$ theory and not only as an $stu$ truncation.
The connection with $N=8$ arises through the formalism of \cite{Ferrara:1997ci}, where U-duality invariant BPS conditions were given for the electric/magnetic charges, and through \cite{D'Auria:1999fa}, where the  U-invariants were related to the eigenvalues of the central charge.

The general procedure proposed in this paper for finding $W$ consists in the following steps:
\begin{itemize}
	
\item[a)] find $W$ in a simple charge and field configuration.

(For the $st^2$ and $stu$ models, this can be one of the three axion free charge configurations: $(q_0,p^0)$, electric $(p^0,q_i)$ or magnetic $(p^i,q_0)$ configuration as discussed in \cite{Ceresole:2007rq}.
In this case, it is known that $W$ is obtained from $Z$ by a series of sign flips.
In the language of \cite{Ceresole:2007wx}, this corresponds to the class of superpotentials obtained from $|Z|$ by acting on the charges by a constant $S$-type  matrix transformation $Q\rightarrow SQ$.)

\item[b)] Use symmetry properties to reconstruct the seed superpotential and then boost it by a duality transformation to generic charges (the pivot method pioneered in \cite{Sen:1994eb,Cvetic:1996zq} and applied to the $stu$ model in \cite{Cvetic:1995uj,Behrndt:1996hu,Gimon:2007mh}).
Whenever the black hole potential has $n_f$ flat directions this leads to a $n_f$-parameter family of superpotentials $W_{\alpha_r}$, $r = 1,
\ldots, n_f$.

\item[c)] Rewrite $W_{\alpha_r}$ expressing the fields in terms of invariants.
This leads to a superpotential with $n_f$ non-invariant fields $b_r$: $W(b_r, i_1,...,i_5)$.

\item[d)] Integrate out the auxiliary fields by solving $\partial_{b_r} W = 0$ and replacing their value in $W$.
This leads to an expression given solely in terms of invariants $(i_1,\ldots,i_5)$.

\end{itemize}
We will also argue that we can obtain $W$ for a generic symmetric space geometry in $N=2$ theories by looking at the $stu$ model and substituting the invariants of the $stu$ model with generic invariants of the desired model.
This procedure can be further generalized to $N=8$ using the fact that $stu$ is both a subsector of $N=8$ and a model in $N=2$ where we know how to describe $W$ in terms of invariants.

This paper is organized along the following lines: in section 2 we adapt the special geometry of generic $N=2$ theories to the $stu$ model and its contractions and discuss their invariants along the lines of \cite{Cerchiai:2009pi}; in section 3 we provide a detailed application of the general method discussed above to the $st^2$ model, which is the minimal rank model with a moduli space; in section 4 we give the relevant results for the $stu$ and $t^3$ models;  in section 5 we discuss the generalization to $N=8$; finally, section 6 contains some concluding comments.

\section{Special geometry for minimal models of rank $1,2,3$.}

Cubic special K\"ahler geometries in $N=2$, $d=4$ supergravities are a subset of the special K\"ahler geometries describing the $\sigma$-model of the scalar fields in the vector multiplets.
The distinguishing feature is the cubic prepotential function $F(X^\Lambda)$, which can arise in the large volume limit of Calabi-Yau compactifications of Type II superstrings or as reduction of minimal supergravity coupled to vector multiplets in $d=5$.

Using special coordinates $z^i=X^i/X^0=x^i-i \,y^i$ ($i=1,\ldots,n$), cubic special K\"ahler manifolds \cite{Gunaydin:1983bi} are described by a set of constants $d_{ijk}$, defining the holomorphic prepotential
\begin{eqnarray}
F(X) &=&\frac{1}{3!}d_{ijk}\frac{X^iX^jX^k}{X^0}=(X^0)^2f(z)\, ,
\label{prepotf0}\\[3mm]
f(z) &=&\frac{1}{3!}d_{ijk}z^i z^j z^k \, .
\end{eqnarray}
Their K\"ahler potential is then
\begin{equation}
{\rm e}^{-K}=-i\left[(f-{\overline f})+\frac12({\bar z}^{\bar\imath}-z^i)({\partial_i f}+
{\bar {\partial}_{\bar\imath}}{\bar f})\right]=-\frac{i}{3!}d_{ijk}
(z^i-{\bar z}^{\bar\imath})(z^j-{\bar z}^{\bar\jmath})(z^k-{\bar z}^{\bar k})\,  .
\end{equation}
This class of models include in particular all symmetric special K\"ahler spaces $G/H$, as classified in \cite{Cremmer:1984hc}.

Since extremal black hole solutions are obtained by minimizing the black hole scalar potential $V_{BH}$ in an effective 1-dimensional action and since $V_{BH}$ is defined in terms of the covariantly holomorphic central charge $Z=e^{K/2}(q_\Lambda X^\Lambda-p^\Lambda F_\Lambda)$, we summarize here some relevant identities that are going to be used throughout this paper:
\begin{eqnarray}
Z &=& e^{K/2}\left[q_0+q_iz^i+p^0f(z)-p^i\partial_i f (z)\right],\\[2mm]
D_i D_j Z &=& i C_{ijk}\, g^{k{\bar k}}\, {\overline {D}}_{\bar k}
{\overline {Z}},\\[2mm]
D_i D_{{\bar\jmath}}{\overline {Z}}&=& g_{i{\bar\jmath}}\, {\overline
{Z}},\label{identities}\\[2mm]
\overline{D}_{\bar\imath} C_{ijk}&=&0,
\end{eqnarray}
where $C_{ijk}$ is a useful tensor defining also the curvature:
\begin{equation}
R_{i{\bar\jmath} k{\bar l}}=-g_{i{\bar\jmath}}\,g_{k{\bar l}}-
g_{i{\bar l}}\,g_{k{\bar\jmath}}+C_{ikp}\,C_{{\bar\jmath}{\bar l}{\bar p}}\,g^{p{\bar p}}. \label{curvature}
\end{equation}
When the moduli space is a cubic symmetric space, one has the additional properties
\begin{eqnarray}
&&\overline{D}_{\bar l} C_{ijk}= 0, \qquad \Rightarrow \quad C_{ijk}=e^K\,d_{ijk},\\[2mm]
&&C_{j(lm} C_{pq)k}\, {\overline C}_{\bar\imath\bar\jmath{\bar k}}
\, g^{j \bar\jmath}\, g^{k{\bar k}}=\frac43\, C_{(lmp}g_{q){\bar \imath}}.
\end{eqnarray}

\paragraph{Invariants in $N=2$ special geometry.}

Following \cite{Cerchiai:2009pi}, we know that a complete set of duality invariants for a given symmetric special geometry is given by the following combinations:
\begin{eqnarray}
i_1 &=& Z{\overline {Z}},\\[2mm]
i_2 &=& g^{i{\bar\jmath}}
Z_i{\overline {Z}}_{\bar\jmath}\,\qquad\qquad (Z_i=D_iZ\,,\
{\overline {Z}}_{\bar\imath}={\overline {D}}_{\bar\imath}\,
{\overline {Z}} )\, ,
\label{i1i2}\\[2mm]
i_3&=&\frac16\left[ Z N_3({\overline {Z}})+{\overline {Z}}
{\overline N}_3 (Z)\right], \label{i3}\\[2mm]
i_4 &=& \frac{i}{6}\left[ Z N_3({\overline {Z}}) -
{\overline {Z}} {\overline N}_3(Z)\right]\,, \label{i4}\\[2mm]
i_5 &=& g^{i{\bar\imath}}C_{ijk}C_{{\bar\imath}{\bar\jmath}\bar k}
{\overline {Z}}^j{\overline {Z}}^k\, Z^{\bar\jmath} Z^{\bar k} \, , \label{i5}
\end{eqnarray}
where
\begin{equation}
N_3({\overline {Z}})=C_{ijk}\, {\overline {Z}}^i\ {\overline {Z}}^j\
{\overline {Z}}^k\,,\qquad \ \qquad {\overline N}_3(Z)=C_{{\bar\imath} {\bar\jmath}\bar k}\, Z^{\bar\imath}\ Z^{\bar\jmath}\ Z^{\bar k} .
\end{equation}
However, these $5$ invariants are not independent, as there is one relation among them, which involves the quartic invariant $I_4$ of symmetric special geometry:
\begin{equation}
I_4=(i_1-i_2)^2+4\, i_4-i_5\, .
\label{I4general}
\end{equation}
Remarkably $I_4$ depends only on the charges and not on the scalar fields, which eventually drop out in the combination (\ref{I4general}).
Note that for all (rank three) symmetric spaces, at the $Z \neq 0$ non-BPS attractor points
\begin{equation}
	i_2 = 3 i_1, \quad i_3 =0, \quad i_4 = -2 i_1^2, \quad i_5 = 12 i_1^2
\end{equation}
and then 
\begin{equation}
	I_4 = - 16 i_1^2 <0.
\end{equation}

Although we will always call such quantities ``duality invariants'', we should stress that they are really scalar functions of the scalar fields and charges under duality transformations.
This means that they will mantain their functional form in terms of the transformed charges and scalar fields, although the specific expression may depend on the details of the case under study.

\paragraph{Minimal models of rank 1, 2 and 3.}

Since we are going to focus on the {\it stu} model and its contractions, it is useful to specialize the identities outlined above to this case.
For the $stu$ model the curvature cannot have mixed indices, because of the factorized structure of the scalar manifold.
Moreover, the triple intersection numbers $C_{ijk}$ are not vanishing only if $i\neq j\neq k$.
This results in a constraint for the product of $C$-tensors
\begin{equation}
	C_{stu}{\bar C}_{{\bar s} {\bar t} {\bar u}}=g_{s\bar s}\, g_{t\bar t}\, g_{u\bar u}
\end{equation}
and in simplified relations for the double covariant derivatives of the central charge:
\begin{eqnarray}
D_s D_t Z=i C_{stu}\, g^{u\bar u}{\bar D}_{\bar u} {\bar Z}\, , &\qquad& D_s D_s Z=0, \\[2mm]
D_s {\bar D}_{\bar s}{\bar Z}=g_{s\bar s}{\bar Z},  &\qquad& D_s {\bar D}_{\bar t} {\bar Z}= D_s {\bar D}_{\bar u}{\bar Z}=0,
\end{eqnarray}
with similar relations for $s\to t\to u$.
These simplified relations are especially useful when computing derivatives of the scalar potential $V_{BH} = i_1 + i_2$ and of the other duality invariant quantities.
Given the factorized structure of the manifold, we can actually split the second invariant into three pieces:
\begin{equation}
	i_2 = g^{i \bar \jmath} D_i Z \overline D_{\bar \jmath}\overline Z = |D_sZ|^2+|D_tZ|^2+|D_uZ|^2 \equiv i_2^s + i_2^t + i_2^u,
\end{equation}
whose derivatives satisfy some interesting relations
\begin{eqnarray}
D_s i_2^s &=& D_s Z{\bar Z}=D_s(i_1), \\[2mm]
D_s i_2^t &=& iC_{stu}g^{u{\bar u}}{\bar D}_{\bar u} {\bar Z}{\bar
D}_{\bar t}{\bar Z}g^{t\bar t},\\[2mm]
D_s i_2^u &=& iC_{stu}g^{t{\bar t}}{\bar D}_{\bar t} {\bar Z}{\bar D}_{\bar u}{\bar Z}g^{u\bar u}.
\end{eqnarray}
It is now obvious that there are special linear combinations that do not depend on some of the moduli.
For instance, $i_2^s - i_1$ and $i_2^t-i_2^u$ are $s$-independent, $i_2^t - i_1$ and $i_2^u-i_2^s$ are $t$-independent and $i_2^u - i_1$ and $i_2^u-i_2^s$ are $u$-independent.
This eventually leads to combinations depending on a single modulus.
The combination
\begin{equation}
i_1 - i_2^s - i_2^t + i_2^u  = Z\bar Z-|D_sZ|^2-|D_tZ|^2+|D_uZ|^2\,
\end{equation}
depends only on $u$ and, similarly, there is a combination depending only on $s$ and one depending only on $t$ that can be obtained by permutations of $s\to t \to u$.

For the $st^2$ model, using again the identities outlined above, one finds a similar result, namely
\begin{equation}
D_t (i_2^t-i_2^s-i_1)=0,\qquad  \qquad D_s(i_2^s-i_1)=0,
\end{equation}
so $i_2^s-i_1$ depends only on $t$ and $i_2^t-i_2^s-i_1$ depends only on $s$.

\section{The $st^2$ model} 

The $st^2$ model is a $\sigma$-model described by the coset manifold $\left[{\rm SU}(1,1)/{\rm U}(1)\right]^2$ with a cubic prepotential
\begin{equation}
	F(X) = \frac{X^1 (X^2)^2}{X^0},
\end{equation}
which falls in the general classification given in (\ref{prepotf0}) for $d_{122} = 2$.
The name of the model is a consequence of the expression of the prepotential in terms of the special coordinates:
\begin{equation}
	s = \frac{X^1}{X^0} \qquad \hbox{and} \qquad t = \frac{X^2}{X^0},
\end{equation}
which leads to $F(X)/(X^0)^2 = f(s,t) = s t^2$.
The K\"ahler potential of this model is
\begin{equation}
	K = - \log\left[-i(s - \bar s)(t - \bar t)^2\right]
\end{equation}
and the central charge governing the BPS flows and defining the black hole potential reads
\begin{equation}
	Z = {\rm e}^{K/2} \left(q_0 + q_1 s + q_2 t -2 p^2 s t - p^1 t^2 + p^0 s t^2\right).
\end{equation}

Following the strategy outlined in the introduction, we are now going to obtain the general ``fake'' superpotential $W$ driving the first order flows for non-BPS extremal black holes.
We start from a simple charge configuration that is known to allow for axion-free truncations.
There are two obvious such configurations, named electric and magnetic depending on which of the two types of charges appear mostly in the definition of the central charge.
We choose to start our construction from the magnetic configuration, with charges
\begin{equation}
	Q = \left(\begin{array}{c}
	0 \\
	P^1 \\
	P^2 \\
	Q_0 \\
	0 \\
	0
	\end{array}\right),
	\label{magcharges}
\end{equation}
though one could obtain the same results by starting from the electric one, which has complementary charges: $Q = (P^0, 0, 0, 0, Q_1, Q_2)$.
We also assume that the signs of the charges allow for non-BPS $Z \neq 0$ black holes.
This implies a negative definite quartic invariant, whose general expression is
\begin{equation}
I_4 = 4q_0p^1(p^2)^2 - p^0q_1(q_2)^2- (p^{0}q_{0}+p^{1}q_{1}+p^{2}q_{2})^2  + 4 p^1q_1 p^2q_2+(p^2q_2)^2.
\end{equation}
For the magnetic configuration discussed here, this means
\begin{equation}
	I_4 = 4 Q_0 P^1 (P^2)^2 <0.
	\label{I4magnetic}
\end{equation}
When all the axions are set to zero, the central charge reduces to a very simple expression:
\begin{equation}
	Z = {\rm e}^{K/2} \left(Q_0 +2 P^2\, y_s y_t +  P^1\, y_t^2 \right),
	\label{Zst20ax}
\end{equation}
where we defined $s \equiv x_s - i \,y_s$ and $t \equiv x_t - i \, y_t$, so that the K\"ahler cone is defined by $y_s >0$, $y_t >0$.
For this configuration we can now obtain a ``fake'' superpotential by applying a simple strategy outlined in \cite{Ceresole:2007wx}, namely to consider invariances of the black hole potential
\begin{equation}
	V_{BH} = -\frac12\,Q^T {\cal M} Q
\end{equation}
by constant charge rotations $Q \to S Q$.
For the magnetic configuration discussed here and with all axions set to zero, the matrix ${\cal M}$ defining the black hole potential $V_{BH}$ reduces to a diagonal form
\begin{equation}
	{\cal M} = { \rm diag} \left\{\frac{y_sy_t^2}{2},\,\frac{y_t^2}{2y_s},\,y_s,\,\frac{1}{2y_sy_t^2},\,\frac{y_s}{2y_t^2},\,\frac{1}{4y_s}\right\}
\end{equation}
and therefore it remains constant under a sign change of any of the charges.
Since the important invariant combination of charges is (\ref{I4magnetic}), we can always redefine the charges so that the only relevant sign change is in front of the electric charge $Q_0$.
This leads to the ``fake'' superpotential
\begin{equation}
	W = {\rm e}^{K/2} \left(-Q_0 +2 P^2\, y_s y_t + P^1\, y_t^2 \right),
	\label{Wst20ax}
\end{equation}
which indeed gives the same black hole potential as the central charge (\ref{Zst20ax}) and produces the critical point $y_s = \sqrt{-P^1 Q_0}/|P^2|$, $y_t = \sqrt{-Q_0/P^1}$.

This ``fake'' superpotential can now be used to generate other generic charge configurations by a duality rotation, following the original idea of \cite{Sen:1994eb} and subsequent developments \cite{Cvetic:1995uj,Cvetic:1996zq,Behrndt:1996hu,Gimon:2007mh}.
However, in order to fully cover the whole orbit of $Z \neq 0$ non-BPS black holes, the seed superpotential needs at least one more parameter and therefore we have to turn on again the axions in (\ref{Wst20ax}).
Since we know that the axion-free $W$ written above and the solutions of the related first order flows are consistent truncations of the general magnetic configuration we can argue that the axions have to appear in (\ref{Wst20ax}) at least quadratically.
The simplest guess is to promote the real fields appearing in (\ref{Wst20ax}) to the full complex combinations $i\, y_s \to s$, $i\, y_t \to t$, or their complex conjugate.
Possible ambiguities are then fixed by the requirement that $W$ be a real function.
Combining these two different sources of information we can propose the following superpotential
\begin{equation}
	W = {\rm e}^{K/2} \left(-Q_0 + P^2\, (s \bar{t} + \bar s t) +  P^1\, t \bar t \right),
	\label{Wst20}
\end{equation}
which indeed is a good generating function for the non-BPS flows in the magnetic configuration, i.e.~it satisfies (\ref{VBH}) and its critical point is a non-BPS critical point of the corresponding $V_{BH}$, for the given charges.

The generic configuration now follows by the action of an SU(1,1)$^2$ duality action on the scalar fields.
This action can be easily obtained by looking at the isometries of the scalar manifold.
Using once more special coordinates, the holomorphic Killing vectors generating the 6 isometries are
\begin{equation}
	k = \left(\begin{array}{c}
	k^s \\ k^t
	\end{array}\right) = \left(\begin{array}{c} (\theta_s-\phi_s) + 2 \psi_s \,s + (\theta_s-\phi_s) s^2 \\
	 (\theta_t+\phi_t) + 2 \psi_t \,t + (\theta_t-\phi_t) t^2	\end{array}\right).
\end{equation}
Here $\theta_s$ and $\theta_t$ denote the 2 compact generators.
As it is known \cite{Strominger:1990pd}, because of general properties of the special K\"ahler geometry nature of the vector multiplet scalar manifolds, these Killing vectors induce a symplectic action on the symplectic sections $V = \{X^\Lambda, F_\Lambda\}$ according to
\begin{equation}
	k_\Lambda^i \partial_i V = T_\Lambda V + f_\Lambda V,
\end{equation}
where $T_\Lambda$ are the 6 corresponding symplectic generators
\begin{equation}
	T =  \left(\begin{array}{cccccc}
	 - \psi_s - 2 \psi_t& - \theta_s+\phi_s & -2 (\theta_t-\phi_t)& 0& 0& 0\\
	  \theta_t-\phi_t & \psi_s - 2 \psi_t& 0& 0& 0& -2 (\theta_t-\phi_t)\\
	  \theta_t+\phi_t & 0& -\psi_s& 0& -2 (\theta_t-\phi_t)& -\theta_s+\phi_s\\
	 0& 0& 0& \psi_s + 2 \psi_t& -\theta_t+\phi_t& -\theta_t-\phi_t\\
	 0& 0& \theta_t+\phi_t& \theta_s - \phi_s& - \psi_s  + 2 \psi_t& 0\\
	 0& \theta_t+\phi_t& \theta_t-\phi_t& 2 (\theta_t-\phi_t)& 0&  \psi_s
	\end{array}\right)
	\label{Tgen}
\end{equation}
and $f_\Lambda$ are the holomorphic functions generating K\"ahler transformations of the K\"ahler potential induced by the same isometries.

The finite action on the scalar fields and charges can be obtained by proper exponentials of the generators.
It is useful to encode it in two SU(1,1) matrices
\begin{equation}
	M_s = \left(\begin{array}{cc}
	a_s & b_s \\
	c_s & d_s
	\end{array}\right), \qquad
	M_t = \left(\begin{array}{cc}
	a_t & b_t \\
	c_t & d_t
	\end{array}\right),
\end{equation}
where
\begin{eqnarray}
	a &=& {\rm e}^{\psi}(\cos \theta \cosh \phi + \sin \theta \sinh \phi), \\
	b &=& {\rm e}^{-\psi}(\sin \theta \cosh \phi + \cos \theta \sinh \phi), \\
	c &=& {\rm e}^{\psi}(-\sin \theta \cosh \phi + \cos \theta \sinh \phi), \\
	d &=& {\rm e}^{-\psi}(\cos \theta \cosh \phi - \sin \theta \sinh \phi),
\end{eqnarray}
so that $a d - bc =1$.
The resulting action on the fields is a fractional transformation with the same parameters:
\begin{equation}
	s \to \frac{a_s \, s + b_s}{c_s\,s+ d_s}, \qquad \qquad t \to \frac{a_t\, t + b_t}{c_t \, t + d_t}.
\end{equation}
As explained above, the same duality transformation must also act on the charges vector $Q$ by rotating it with a symplectic rotation $Q \to S Q$, where
\begin{equation}
	S_s = \left(\begin{array}{cccccc}
	 d_s&c_s&&&& \\
	 b_s&a_s&&&& \\
	 &&d_s&&& \frac12\,c_s\\
	 &&&a_s&-b_s& \\
	 &&&-c_s&d_s& \\
	 &&2b_s&&&a_s
\end{array}\right)
\end{equation}
is the generator induced by the SU(1,1) factor acting only on the $s$ field and
\begin{equation}
	S_t = \left(\begin{array}{cccccc}
	 d_t^2&&2 c_t d_t&&c_t^2& \\
	 &d_t^2&&-c_t^2&& c_t d_t\\
	 b_t d_t&&a_t d_t + b_t c_t&&a_t c_t& \\
	 &- b_t^2&&a_t^2&& -a_t b_t\\
	  b_t^2&&2a_t b_t&&a_t^2& \\
	 &2b_t d_t&&-2 a_t c_t&&a_t d_t + b_t c_t
	\end{array}\right)
\end{equation}
is the SU(1,1) action induced by the action on the $t$ field.
Both can be obtained by proper combinations of the exponentials of the symplectic generators (\ref{Tgen}).
At this point we can write explicitly the transformation mapping the magnetic configuration of charges discussed above, starting with (\ref{magcharges}), to the generic configuration parameterized by $Q^\prime = (p^0,p^1,p^2,q_0,q_1,q_2)$.
Solving the linear system $S_s S_t Q = Q^\prime$ fixes the values of the coefficients of the fractional transformations to
\begin{equation}
	M_s =\frac{sgn(\nu_s)}{\sqrt{2(\sigma^{+}_s+\sigma^{-}_s)\rho_s
	\nu_s}}
\left(\begin{array}{cc}
\sigma^{+}_s \nu_s+\sigma^{-}_s & \rho_s(\sigma^{+}_s \nu_s-\sigma^{-}_s) \\
\nu_s-1 & \rho_s(\nu_s+1)
\end{array}\right),
\end{equation}
\begin{equation}
	M_t =\frac{sgn(\nu_t)}{\sqrt{2(\sigma^{+}_t+\sigma^{-}_t)\rho_t
	\nu_t}}
\left(\begin{array}{cc}
\sigma^{+}_t \nu_t+\sigma^{-}_t & \rho_t(\sigma^{+}_t \nu_t-\sigma^{-}_t) \\
\nu_t-1 & \rho_t(\nu_t+1)
\end{array}\right),
\end{equation}
where
\begin{equation}
	\rho_s=\frac{\sqrt{-Q_0 P^1}}{P^2}, \qquad
	\rho_t=\sqrt{-\frac{Q_0}{P^1}},
\end{equation}
given in terms of the original charges, and
\begin{equation}
	\sigma^{\pm}_s=\frac{\sqrt{-I_4}\pm \left(p^{\Sigma}q_{\Sigma}-2p^1 q_1\right)}{2(p^2)^2-2 p^0q_1}, \qquad 	
	\sigma^{\pm}_t=\frac{\sqrt{-I_4}\pm \left(p^{\Sigma}q_{\Sigma}-2p^2q_2\right)}{2p^1p^2-2 p^0q_2} \label{sigma},
\end{equation}
and
\begin{equation}
	\nu_s = \alpha^2 \nu, \qquad \nu_t = \frac{1}{\alpha} \nu,
	\label{nudef}
\end{equation}
with
\begin{equation}
\nu^3= \nu_s \nu_t^2=\frac{2 p^1(p^2)^2+p^0(\sqrt{-I_4}- p^{\Sigma}q_{\Sigma})} {2 p^1(p^2)^2-p^0(\sqrt{-I_4}+p^{\Sigma}q_{\Sigma})},
\end{equation}
given in terms of the final charges.

The outcome of this procedure is
\begin{equation}
	W_\alpha = \frac{(-I_4)^{1/4}}{4\, \sqrt{Y^1}\, Y^2} \left[Y^1\,(Y^2)^2+Y^1\,(X^2)^2+2X^1\,X^2\,Y^2+Y^1+2Y^2\right],
	\label{wst2}
\end{equation}
where we have introduced 4 new real combinations of the coordinates to simplify the expression of the ``fake'' superpotential:
\begin{eqnarray}
	X^i &=&\frac{\nu_i^2|\sigma^{+}_i-z^i|^2-|\sigma^{-}_i+z^i|^2}{i(z^i- \bar{z}^i)\,
	\nu_i(\sigma^{+}_i+\sigma^{-}_i)},
	\label{Xcoor}\\[2mm]
	Y^i &=&\frac{|\nu_i\sigma^{+}_i+\sigma^{-}_i+(1-\nu_i)z^i|^2}{i(z^i- \bar{z}^i)\, \nu_i(\sigma^{+}_i+\sigma^{-}_i)}
	\label{Ycoor}.
\end{eqnarray}
It is a tedious, but straightforward exercise to check that (\ref{wst2}) reproduces the full black hole potential using (\ref{VBH}).

An important feature of the function $W_\alpha$ we have derived so far is given by the fact that everything is determined in terms of the charges with the exception of the coefficient $\alpha$.
This means that we actually obtained a one-parameter family of candidate ``fake'' superpotentials giving the same black hole potential.
The reason for this can be explained by looking at the first order flows driven by $W_\alpha$ and at its critical points.
Setting $\partial_s W_\alpha = 0$ and $\partial_t W_\alpha =0$ one obtains that the critical points lie at
\begin{equation}
	s = \frac{\sigma^{+}_s\nu_s^2-\sigma^{-}_s- i(\sigma^{+}_s+\sigma^{-}_s)\nu_s}	{\nu_s^2+1}, \qquad \qquad
	t=\frac{\sigma^{+}_t\nu_t^2-\sigma^{-}_t- i (\sigma^{+}_t+\sigma^{-}_t)\nu_t}	{\nu_t^2+1}.
	\label{critalpha}
\end{equation}
The specific value of these critical points depend on the charges, contained in the $\sigma_{s,t}^\pm$ and $\nu$ functions, but also on the real parameter $\alpha$ appearing in the definition of $\nu_s$ and $\nu_t$ in terms of $\nu$ (see (\ref{nudef})).
We therefore see that, for each value of $\alpha$ and hence for each $W_\alpha$, there is a unique, well defined, critical point.
This is different from what happens if one looks at $V_{BH}$.
The full black hole potential has a line of critical points, i.e.~a one-dimensional moduli space for the resulting non-BPS $Z \neq 0$ black hole configurations.
We will now explain why this is possible, how to recover a universal $W$ and how we can turn this feature into a desirable property.

The first thing to notice is that critical points of $W_\alpha$ are not always minima. Although $W_\alpha$ drives first order flows, just like the central charge, and although the non-BPS black hole critical points are given by critical points of $W_\alpha$, just like the BPS ones are given by critical points of $Z$, there is no geometric relation forcing the Hessian to be positive definite.
This is an important difference with respect to the BPS case, where the (covariantly) holomorphic nature of $Z$ and special geometry relations impose that all critical points are minima.
While the central charge $Z$ will give well defined basins of attractions, $W_\alpha$ may also have parts of the moduli space that are repelled from the critical point.
Starting from one of these points in moduli space will lead to the so-called ``flows to hades'', i.e.~solutions leading to naked singularities.
This feature is related to the existence of a non-trivial moduli space at the black hole horizon and it is actually a necessary property for a consistent $W_\alpha$.
Since $W_\alpha$ has a unique critical point for each different value of $\alpha \in {\mathbb R}$, one per critical point in the black hole moduli space, all the other critical points of $V_{BH}$ can be used as asymptotic values of the moduli for $W_\alpha$ with a non-trivial gradient flow.
This obviously is in sharp contrast with the expectation that these points in moduli space are critical points where no black hole flow should start.
However, if they lie in a basin of repulsion of $W_\alpha$ they will never lead to black hole solutions and therefore all black holes described by $W_\alpha$ for fixed $\alpha$ are a subset of those described by $V_{BH}$.
Varying the value of $\alpha$, one then obtains all the non-BPS solutions contained in $V_{BH}$.
We can show this explicitly by solving the non-BPS flow analytically.

Using the redefined coordinates introduced in (\ref{Xcoor})--(\ref{Ycoor}), the flow equations have the form
\begin{eqnarray}
&&\!\!\!\!\!\!\!\!\!\!\!\!\!\!\!U^\prime=- e^{U} \,{W}= - e^{U}\frac{(-I_4)^{1/4}}{4\, \sqrt{Y^1}\, Y^2} \left[Y^1\,(1+(X^2)^2+(Y^2)^2)+2Y^2(1+X^1\,X^2)\right], \label{warp}\\[3mm]
&&\!\!\!\!\!\!\!\!\!\!\!\!\!\!\!\frac{{X^1}^\prime}{X^1}=-e^{U}\frac{(-I_4)^{1/4}}{2\sqrt{Y^1} Y^2}\left[Y^1\left(1+(X^2)^2+(Y^2)^2\right)+4\frac{Y^2 X^2}{X^1}+ 2 Y^2\left(-1+X^1\,X^2\right)\right],
\label{X1prime}\\[3mm]
&&\!\!\!\!\!\!\!\!\!\!\!\!\!\!\!\frac{{X^2}^\prime}{X^2}=-e^{U}\frac{(-I_4)^{1/4}}{2\, \sqrt{Y^1}\, Y^2}\left[Y^1\,\left(1+(X^2)^2+(Y^2)^2\right)+2\frac{Y^2}{X^2}\,\left(X^1+(X^2)^2\,X^1\right)\right],
\label{X2prime}\\[3mm]
&&\!\!\!\!\!\!\!\!\!\!\!\!\!\!\!\frac{{Y^1}^\prime}{Y^1}=-e^{U}\frac{(-I_4)^{1/4}}{2\, \sqrt{Y^1}\, Y^2}\left[Y^1\,\left(1+(X^2)^2+(Y^2)^2\right)+2Y^2\,\left(-1+X^2\,X^1\right)\right],
\label{Y1prime}\\[3mm]
&&\!\!\!\!\!\!\!\!\!\!\!\!\!\!\!\frac{{Y^2}^\prime}{Y^2}=-e^{U}\frac{(-I_4)^{1/4}}{2\, \sqrt{Y^1}\, Y^2}\left[Y^1\,\left(-1+(X^2)^2+(Y^2)^2\right)+2Y^2\,X^2\,X^1\right].
\label{Y2prime}
\end{eqnarray}
If we first consider the difference between (\ref{warp}) and (\ref{Y1prime}), with a factor $1/2$, we obtain a new equation depending only on the warp factor and $Y^1$:
\begin{equation}
({\rm e}^{-U}\sqrt{Y^1})^\prime=(-I_4)^{1/4}.
\end{equation}
This can be easily solved by
\begin{equation}
	{\rm e}^{-2 U} Y^1 = h_2^2,
	\label{solY1}
\end{equation}
where $h_2=c_2+(-I_4)^{1/4}\tau$ is a harmonic function and $c_2$ is a real integration constant.
A similar relation can also be obtained for $Y^2$, by taking the difference between (\ref{Y1prime}) and (\ref{Y2prime}):
\begin{equation}
	\frac{Y^1{}^\prime}{Y^1}- \frac{Y^2{}^\prime}{Y^2} = -{\rm e}^U (-I_4)^{1/4}\frac{\sqrt{Y^1}}{Y^2}+{\rm e}^U(-I_4)^{1/4}\frac{1}{\sqrt{Y^1}}.
\end{equation}
Using (\ref{solY1}) in the last term, this equation can be rewritten as
\begin{equation}
	\frac{Y^1{}^\prime}{2 Y^1} + U^\prime - \frac{Y^2{}^\prime}{Y^2}=-(-I_4)^{1/4} \left(\frac{{\rm e}^U \sqrt{Y^1}}{Y^2}\right).
\end{equation}
We can easily also rewrite the left hand side as
\begin{equation}
	\frac{Y^1{}^\prime}{2 Y^1} + U^\prime - \frac{Y^2{}^\prime}{Y^2} = \left(\frac{{\rm e}^U \sqrt{Y^1}}{Y^2}\right)^{-1} \left(\frac{{\rm e}^U \sqrt{Y^1}}{Y^2}\right)^\prime
\end{equation}
and hence integrate the equation to
\begin{equation}
	\frac{Y^2}{{\rm e}^U \sqrt{Y^1}} = h_1,
\end{equation}
where now $h_1=c_1+(-I_4)^{1/4}\tau$ is another harmonic function.
Using again (\ref{solY1}) we eventually get that
\begin{equation}
{\rm e}^{-2 U} \, Y^2=\, h_2\, h_1.
\end{equation}
Another similar pattern can be applied to integrate the equations for the axion combinations $Z^i \equiv {\rm e}^{-2 U} X^i$.
The relevant linear combinations can be obtained by looking at the difference between (\ref{warp}) and half of (\ref{X1prime}) and at the difference of (\ref{warp}) and half of (\ref{X2prime}).
The flow equations become
\begin{eqnarray}
Z^1{}^\prime=2\frac{(-I_4)^{1/4}}{h_2}(Z^1-Z^2),
\label{Z1prime}\\[2mm]
Z^2{}^\prime=\frac{(-I_4)^{1/4}}{h_2}(Z^2-Z^1),
\label{Z2prime}
\end{eqnarray}
which also leads to
\begin{equation}
	(Z^2-Z^1)^\prime=-\frac{(-I_4)^{1/4}}{h_2}(Z^2-Z^1). \label{rela}
\end{equation}
From the flow equations one also gets the following relation
\begin{equation}
\frac{{(X^2-X^1)^\prime}}{X^2-X^1} +\frac{{(Y^2-Y^1)^\prime}}{Y^2-Y^1}-2 \frac{{Y^1}^\prime}{Y^1}=0,
\end{equation}
which, by plugging in factors of ${\rm e}^{-2U}$ and using previous results, can also be rewritten as
\begin{equation}
\frac{(Z^2-Z^1)^\prime}{Z^2-Z^1}+\frac{(h_2(h_2-h_1))^\prime}{h_2(h_2-h_1)}-2 \frac{(h_2^2)^\prime}{h_2^2}=0.
\end{equation}
Using (\ref{rela}) this leads to an inconsistency unless $Z_2 = Z_1$.
This further implies that the rescaled axions $Z^i$ are constants and that therefore
\begin{equation}
X^1=X^2=c\,e^{2U}.
\label{Xcond}
\end{equation}
Substituting the solutions for $X^i$ and $Y^i$ back into the warp factor equation we obtain that
\begin{equation}
{\rm e}^{-4U}=h_0\,h_1\,h_2^2-c^2,
\end{equation}
where $h_0=c_0+(-I_4)^{1/4}\tau$ is another harmonic function.
Although we have introduced 4 different integration constants, only 3 of them are independent because of the constraint coming from the requirement that our metric asymptotes Minkowski spacetime at infinity: $c^2=c_0\,c_1\,c_2^2-1$.
This solution implicitly depends on the $\alpha$ parameter, because it enters in the definition of $X^i$ and $Y^i$ and therefore in the relation between $s$, $t$ and the harmonic functions obtained here.
The integration constants will depend on the asymptotic values of the moduli fields, which in turn will depend on this parameter $\alpha$.
Solutions which start from the repulsion region are going to have a value of the integration constants such that at some point along the flow ${\rm e}^{-4U}$ will vanish, i.e.~the flow will hit a singular point and hence it will not generate a consistent black hole solution.

\begin{figure}[ht]
	\centering
		\includegraphics[scale=.83]{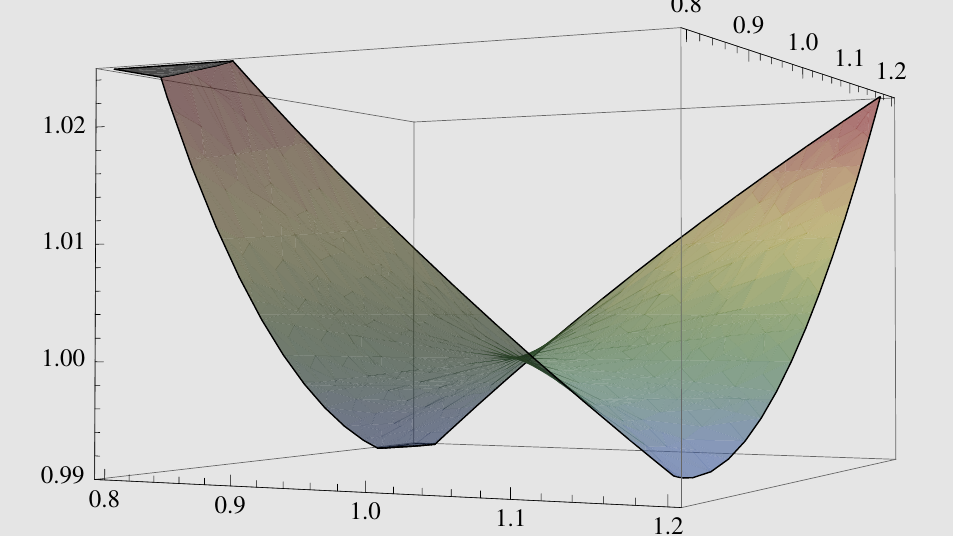}\quad
		\includegraphics[scale=.83]{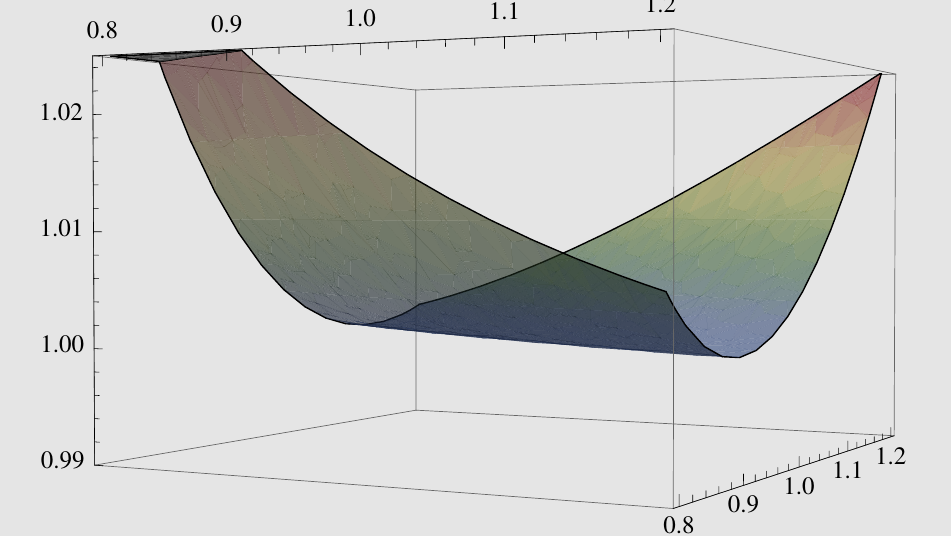}
\caption{\small The plot on the left is a section of the potential $W_\alpha$ for a configuration of unit $q_0$, $p^0$ charges and zero axions, for the value $\alpha =1$. The only critical point is a clear saddle point. The plot on the right is the ``fake'' superpotential $W$ obtained by integrating out $\alpha$, for the same charge configuration and vanishing axions. The unstable direction is now a flat direction at the bottom of the potential, which has a unique basin of attraction.}
	\label{figura}
\end{figure}

For the same given set of charges, we can now produce a universal ``fake'' superpotential where the unstable directions are replaced by flat directions corresponding to the moduli space of the non-BPS black hole attractor.
Since the parameter $\alpha$ is a real constant, we can think of it as a Lagrange multiplier enforcing the constrained variation leading to the black hole flows towards the critical point (\ref{critalpha}).
Black hole solutions are then described by flows where $\partial_\alpha W = 0$.
This is a quartic equation in $\alpha$, which has only one physical solution.
For instance, in a setup where only $p^0$ and $q_0$ are not vanishing, the $\partial_\alpha W = 0$ constraint reads
\begin{equation}
	\alpha^4 (p^0)^{2/3}\,  s \bar s (t-\bar t)	+q_0^{2/3} \alpha^3 (s-\bar s) - \alpha (p^0)^{2/3} t \bar t (s- \bar s) - q_0^{2/3} (t - \bar t)  =0.
\end{equation}
The physical solution can be identified as the only root that reproduces the known critical points of $V_{BH}$ when its expression is plugged back in $W_\alpha$.
In figure~\ref{figura} we show this mechanism for a simple configuration.

As we discussed in the introduction, a proper ``fake'' superpotential has to be invariant under duality transformations along the flow \cite{Andrianopoli:2009je}, and this is true for both $W_\alpha$ as well as $W$.
However, we can now show that the superpotential that includes also the information about the flat direction is an actual scalar under duality transformations on the whole moduli space.
The procedure to show this is the following.
We first rewrite $W_\alpha$ in terms of invariants.
Since this will be only possible in the region of the moduli space leading to proper black hole solutions, it will be done by rewriting 3 out of the 4 real scalar fields in terms of the duality invariants described in the previous sections and leaving out one combination of the scalar fields (which includes the parameter $\alpha$), which is parameterizing the direction orthogonal to the black hole flows and which is not invariant under the same duality transformations.
Since this new combination does not appear in the black hole potential, it can be thought of as an auxiliary field, constant along the flow, which we can therefore integrate out in order to obtain a fully invariant W.
We begin by identifying the 3 invariant combinations of the scalar fields that appear in both the superpotential and the full scalar potential.
Again, this can be best done in a simple charge configuration.
Once these combinations are expressed in terms of the independent invariants, they are valid for any configuration.
In this case, a useful setup is given by the configurations with only $p^0$ and $q_0$ not vanishing.
The superpotential obviously depends on all the scalar fields, but the scalar potential depends only on the overall volume modulus and not on the ratio between $y_s$ and $y_t$.
By using some properties of the invariants noted in the previous section, namely the fact that there are combinations that are independent on some of the scalar fields, we can parameterize the axions and the overall volume modulus in terms of invariants as follows
\begin{eqnarray}
	a_1&=&\frac{i_1+i_2^{s}-i_2^{t}}{\sqrt{-I_4}} =- \frac{x_s}{y_s},\\[2mm]
	a_2&=&\frac{i_1-i_2^{s}}{\sqrt{-I_4}} =- \frac{x_t}{y_t},\\[2mm]
	m&=&\frac{-I_4 (i_1+i_2^{s}+i_2^{t})+a_1 a_2^2(-I_4)^{3/2}+4 i_3\sqrt{-I_4}}{(-I_4)^{3/2}} = \frac{q_0}{p^0 y_s y_t^2}.
\end{eqnarray}
Although we derived these expressions in a special duality frame (the last equality of each line is valid only when $p^0$ and $q_0$ are different from zero with all the other charges vanishing), the results we presented can be trusted in any other frame because of their invariant expression.

Using these invariant combinations of the coordinates, we can rewrite the coordinates introduced in (\ref{Xcoor}) and (\ref{Ycoor}) as
\begin{equation}
	Y^i=\frac{(b_i m^{1/3}-a_i)^2+1}{2 b_i m^{1/3}}, \qquad \qquad  X^i= \frac{b^2_i m^{2/3}-a_i^2-1}{2 b_i m^{1/3}},
\end{equation}
where $b_1 = 1/b^2$ and $b_2 = b$, and $b$ is a function of the scalar fields that cannot be expressed in terms of the invariant combinations above.
By a quick comparison of this new definition with (\ref{Xcoor}) and (\ref{Ycoor}) we see that this non-invariant combination depends on $\alpha$, as expected.
Although the fake superpotential $W_\alpha$ depends on all four combinations of the scalar fields
\begin{eqnarray}
W_\alpha =\frac{(-I_4)^{1/4}}{2\sqrt{2}}\frac{m^{1/3} \left[-a_1 \left(1+a_2^2\right)+m\right]+2 m^{2/3} b+
\left[\left(1+a_1^2\right) \left(1+a_2^2\right)-a_1 m\right] b^2} {\sqrt{m \left[b^4 + \left(a_1 b^2 -m^{1/3}\right)^2\right]}},
\label{Wa358}
\end{eqnarray}
the black hole potential has a flat direction, parameterized by the combination $b$:
\begin{equation}
	V_{BH}=\frac{\sqrt{-I_4}}{2 m}\left[\left(1+a_1^2\right)
	\left(1+a_2^2\right)^2-2 a_1 a_2^2 m+m^2\right].
\end{equation}
Extrema of the black hole potential are given by
\begin{equation}
m=1, \quad	a_1=a_2 =0,
\end{equation}
while extrema of the superpotential require a further condition fixing the value of $b$, which means that one fixes the ratio of the volume moduli in relation to the chosen value of $\alpha$.
This fact explains why we can think of $b$ as an auxiliary field in the superpotential that can be integrated out, by setting
\begin{equation}
	\partial_b W_{\alpha} = b\left[1+a_2^2-(1+a_1^2)b^3\right]+m^{2/3}\left[1-b^3\right] =0.
\end{equation}
Out of the four roots of this equation, once again, only one is physical and it corresponds to the solution recovering the known critical points.
Substituting this expression for $b$ in $W_\alpha$ we get rid of the dependence on the extra parameters and we obtain a new expression for the superpotential that is given only in terms of the invariants.

The resulting $W$ should match the one derived by using a different approach in \cite{Bossard:2009we}.
Also the result in \cite{Bossard:2009we} is given in terms of duality invariant quantities and it is only implicitly described by an equation of degree 4 for $W^2$.
On the other hand, as we have detailed in this section, by using an auxiliary field we have an explicit expression for $W$ (\ref{Wa358}) and the solution to its flow equations.


\section{The $stu$ model}

The method we have described in detail in the previous section for the derivation of an invariant superpotential can be applied also to the stu model.
The main difference between these two cases is given by the dimension of the moduli space (2 in this case versus 1 in the previous case) and therefore by the number of auxiliary fields involved in the process of writing an explicit ``fake'' superpotential.
Minimization of the superpotential in terms of the auxiliary fields (parameterizing the non-invariant directions in moduli space) gives now a system of polynomial constraints whose solution will lead to the invariant superpotential.
In this section we will not repeat all the steps of the derivations, but we will present the main results and novelties with respect to the previous case.

The $stu$ model was already thoroughly studied in the context of the first order formalism in \cite{Bellucci:2008sv} and we will use its results as a basis for our discussions in this section.
Following the ``pivot'' method outlined before, the authors of \cite{Bellucci:2008sv} constructed a 2-parameter family of superpotentials for this model:
\begin{eqnarray}
W&=&\textstyle\frac{1}{4}\frac{(-I_4)^{1/4}}{ \sqrt{ \prod_i \nu_i(\sigma^{+}_i+\sigma^{-}_i)}}e^{K/2}
\prod_i\left|\nu_i(\sigma^{+}_i-z^i)+\sigma^{-}_i+z^i\right|\cdot
\label{fake}\\[3mm]
&& \textstyle \cdot \left(1+\sum_{i<j} \frac{\left(\nu_i^2|\sigma^{+}_i-z^i|^2-|\sigma^{-}_i+z^i|^2\right)
\left(\nu_j^2|\sigma^{+}_j-z^j|^2-|\sigma^{-}_j+z^j|^2\right)- \nu_i\nu_j(\sigma^{+}_i+\sigma^{-}_i)(\sigma^{+}_j+\sigma^{-}_j)(z^i- \bar{z}^i)
(z^j-\bar{z}^j)} {\left|\nu_i(\sigma^{+}_i-z^i)+\sigma^{-}_i+z^i\right|^2\,\left|
\nu_j(\sigma^{+}_j-z^j)+\sigma^{-}_j+z^j\right|^2}\right),\nonumber
\end{eqnarray}
where we have used definitions similar to the ones of the previous section.
The functions appearing in (\ref{fake}) depend on the scalar fields and charges according to:
\begin{equation}
\sigma^{\pm}_i=\frac{\sqrt{-I_4}\pm \left(p^{\Sigma}q_{\Sigma}-2p^iq_i\right)}{|\varepsilon_{ijk}|p^jp^k-2 p^0q_i} \quad ({\rm no\; sum\; over}
\;i)\label{sigma2}
\end{equation}
and
\begin{equation}
\nu^3=\prod_i \nu_i=\frac{2 p^1p^2p^3+p^0(\sqrt{-I_4}- p^{\Sigma}q_{\Sigma})} {2 p^1p^2p^3-p^0(\sqrt{-I_4}+p^{\Sigma}q_{\Sigma})}.
\end{equation}
Here the index $\Sigma$ runs over $(0,\,1,\,2,\,3)$, $\varepsilon_{ijk}$ is the totally  antisymmetric tensor of rank 3, $\nu_i$ are related to the superpotential parameters $\alpha_i$ by
\begin{equation}
\nu_i=\nu e^{\alpha_i}\label{nu},
\end{equation}
and $I_4$ is a quartic invariant
\begin{equation}
I_4 = 4q_0p^1p^2p^3 - 4p^0q_1q_2q_3- (p^\Sigma q_\Sigma)^2  + 4\sum_{i<j} p^iq_i p^jq_j.
\end{equation}
Since we have a 2-parameter family of superpotentials, the $\alpha_i$ constants have to satisfy a linear condition $\sum_{i}\alpha_i=0$.
We can once more see that the existence of these $\alpha_i$ parameters is related to flat directions in the potential and also that they depend on the asymptotic value of the scalar fields at infinity.
In facts, two different legitimate choices for $\nu_i$ are
\begin{equation}
	\nu_1 = \nu_2 = \nu_3 = \nu = \left(\frac{2 p^1 p^2 p^3 + p^0 \left(\sqrt{-I_4}- p^\Sigma q_\Sigma\right)}{2 p^1 p^2 p^3 - p^0\left(\sqrt{-I_4}+p^\Sigma q_\Sigma\right)}\right)^{1/3},
\end{equation}
when all $\alpha_i =0$, and
\begin{equation}
	\nu_i = \frac{2^{-1/3}\left(|\epsilon_{ijk}|p^j p^k - 2 p^0 q_i\right)}{\left(2 p^1 p^2 p^3 - p^0 \left(\sqrt{-I_4} + p^\Sigma q_\Sigma\right)\right)^{2/3}},
\end{equation}
with some $\alpha_i \neq 0$ and these choices depend on the value of the moduli fields infinitely far away from the horizon.
Following the line of the previous section, it is useful to introduce new coordinates
\begin{eqnarray}
&&X^i =\frac{\nu_i^2|\sigma^{+}_i-z^i|^2-|\sigma^{-}_i+z^i|^2}{i(z^i- \bar{z}^i)
\nu_i(\sigma^{+}_i+\sigma^{-}_i)},\\[2mm]
&&Y^i =\frac{|\nu_i\sigma^{+}_i+\sigma^{-}_i+(1-\nu_i)z^i|^2}{i(z^i- \bar{z}^i) \nu_i(\sigma^{+}_i+\sigma^{-}_i)},
\end{eqnarray}
whose use allows for a compact rewriting of the ``fake'' superpotential (\ref{fake}):
\begin{equation}
{W}=\frac{(-I_4)^{1/4}}{4\, \sqrt{Y^1}\, \sqrt{Y^2} \,\sqrt{Y^3}} \left (Y ^ 1 \,Y
^2\,Y^3+Y^1\,X^2\,X^3+X^1\,Y^2\,X^3+X^1\,X^2\,Y^3+Y^1+Y^2+Y^3\right). \label{w}
\end{equation}
At this stage we can introduce 4 invariant combinations
\begin{eqnarray}
\!\!\!\!a_i=&&\frac{i_1-\sum_i i_{z^i}+2 i_{z^i}}{\sqrt{-I_4}}
=\frac{1-(Y^i)^2+(X^i)^2}{2 Y^i}, \label{ai}\\[2mm]
\!\!\!\!m=&&\frac{-I_4 (i_1+\sum_i i_{z^i})+\left(\prod_i a_i\right) (- I_4)^{3/2}+4 i_3\sqrt{-I_4}}{(-I_4)^{3/2}} =\prod_i\frac{1+ \left(Y^i+ X^i\right)^2}{2
Y^i}, \label{m}
\end{eqnarray}
which can be used to rewrite $W$ in terms of invariants by the inverse relations
\begin{eqnarray}
Y^i=\frac{(b_i m^{1/3}-a_i)^2+1}{2 b_i m^{1/3}}, \qquad \qquad X^i= \frac{b^2_i m^{2/3}-a_i^2-1}{2 b_i m^{1/3}},
\end{eqnarray}
where we had to introduce two new independent functions parameterizing the non-invariant $n_f$ flat directions.
Once more, we did so by introducing $n_f + 1$ scalar fields $b_i$ subject to the condition $\prod_i b_i=1$.
To see that these fields parameterize the flat directions, we can substitute them in the expression of the black hole potential and show that it depends only on the four invariants (\ref{ai})--(\ref{m}), but not on $b_i$.
On the other hand, the ``fake'' superpotential for the $stu$-model depends on all 6 scalar field combinations, as we can see explicitly from its rather long and intricated expression:
\begin{eqnarray}
W&=&\textstyle \frac{\left(-I_4\right)^{1/4}}{2 \sqrt{2} \sqrt{m \left(b_2^2 b_3^2+\left(a_1b_2 b_3-m^{1/3}\right)^2\right)
\left(1+\left(a_2-m^{1/3} b_2\right)^2\right)
\left(1+\left(a_3-m^{1/3} b_3\right)^2\right)}} \cdot\label{WSTU}\\[2mm]
&\cdot&\left\{\left(1+a_1^2\right) \left(1+a_2^2\right) \left(1+a_3^2\right)
b_2 b_3-m^{1/3}\left[a_1 \left(1+a_2^2\right) \left(1+a_3^2\right)\right.\right.\nonumber\\[2mm]
&+&\left.\left.a_2 \left(1+a_1^2\right) \left(1+a_3^2\right) b_2^2 b_3+a_3\left(1+a_1^2\right)
\left(1+a_2^2\right)  b_2 b_3^2\right]
\right.\nonumber\\[2mm]
&+&\left.\left[\left(1+a_3^2\right)(1+a_1 a_2)  b_2+ \left(1+a_2^2\right) (1+a_1 a_3) b_3+\left(1+a_1^2\right) (1+a_2 a_3)
b_2^2 b_3^2\right] m^{2/3}\right.\nonumber \\[2mm]
&-&\left.2b_2 b_3 (a_1+a_2+a_3+a_1 a_2 a_3) m+\left[1+b_2 b_3 (b_2+a_1 a_3 b_2+b_3)\right.\right. \nonumber \\[2mm]
&+&\left.\left.a_2 \left(a_3+a_1 b_2 b_3^2\right)\right]
m^{4/3}-\left(a_3 b_2+a_2 b_3 +a_1 b_2^2 b_3^2\right) m^{5/3}+b_2 b_3 m^2\right\}.\nonumber
\end{eqnarray}

As in the $st^2$ case, $W$ fixes also the auxiliary fields to a specific value in order to remove possible flows not describing black holes from its solutions.
We now have a system of equations following from
\begin{equation}
\partial_{b_r}{W}=0.\label{cond}
\end{equation}
Substituting the solution to this system $b_r = b_r(a_i,m)$ back in the fake superpotential we get again its invariant expression, which explicitly shows 2 flat directions, i.e.~${W}={W}(a_i,\,m,\,b_r(a_i,m))$ has zero determinant of the Hessian.
Unfortunately, for this specific model it is very hard to write out the solutions for the auxiliary fields because the coupled system reduces to polynomial equations of degree 6.

However, in order to check our procedure, we can investigate a special, but well-known case.
It is easy to see from (\ref{w}) that it is self-consistent to set $X^i=0$.
In the language of invariants $(a_i,\,m)$ this fixes the auxiliary fields to
\begin{equation}
b_r=\frac{a_r^2+1}{m^{2/3}}, \qquad \qquad m=\prod_i \sqrt{1+a_i^2}.
\end{equation}
These equations provide a solution to the auxiliary fields constraint (\ref{cond}), but it also minimizes the ``fake'' superpotential in the direction $m$: $\partial_{m}{W}=0$.
This is related to the fact that this particular configuration has
\begin{eqnarray}
i_3=0.
\end{eqnarray}
Therefore, we can see that the ``fake'' superpotential reduces to
\begin{equation}
W=\frac{1}{2}\left(\sqrt{i_1}+\sqrt{i_2^s}+\sqrt{i_2^t}+\sqrt{i_2^u}\right),
\end{equation}
an expression already discussed in \cite{Andrianopoli:2007gt,Bossard:2009we}.

\section{Extension from the {\it stu} model to an arbitrary $N=2$ symmetric coset theory}

The derivation strategy we used in the previous sections to obtain the ``fake'' superpotential driving the non-BPS black hole solutions for the $st^2$ and $stu$ model can obviously be applied to any other model.
However, given the special structure of the $stu$ model and its prominence as an $N=2$ model arising also as a truncation of $N=8$ supergravity we will use this section to argue that its superpotential could also be used to write down the generic superpotential of any other model in this class, by replacing the duality-invariant combinations of the $stu$ model in terms of the generic invariants of other models (at least for coset scalar manifolds).
In order to do so, we will use a procedure similar to that used for the $N=8$ theory in \cite{D'Auria:1999fa}.

We start our discussion by rewriting the invariants of the $stu$ model as
\begin{eqnarray}
i_1 &=& Z{\bar Z}\\[2mm]
i_2 &=& i_2^s+i_2^t+i_2^u\\[2mm]
i_3^2+i_4^2 &=& 4 i_1\, i_2^s\, i_2^t\, i_2^u\\[2mm]
i_5 &=&4( i_2^s\, i_2^t+i_2^s \,i_2^u+i_2^t\, i_2^u)\, .
\end{eqnarray}
These combinations become suggestive on an underlying structure relating them.
By defining also
\begin{equation}
\lambda_1=i_2^s, \quad \quad \lambda_2=i_2^t, \quad \quad \lambda_3=i_2^u \, ,
\end{equation}
we see that the invariant combinations written above resemble the structure of the coefficients of a cubic polynomial, where $\lambda_i$ are its real roots:
\begin{equation}
\Pi_{i=1}^3 (\lambda-\lambda_i)=0\, ,
\end{equation}
that is
\begin{equation}
\lambda^3-(\lambda_1+\lambda_2+\lambda_3)\lambda^2+(\lambda_1 \lambda_2 +\lambda_1 \lambda_3+\lambda_2\lambda_3)\lambda- \lambda_1 \lambda_2
\lambda_3 =0\, .
\label{poly3}
\end{equation}
In facts, the combinations of invariants discussed above in this language become
\begin{eqnarray}
i_2 &=& \lambda_1+\lambda_2 +\lambda_3, \label{def1}\\[2mm]
\frac{i_3^2+i_4^2}{4i_1} &=& \lambda_1 \lambda_2 \lambda_3, \\[2mm]
\frac{i_5}{4} &=& \lambda_1 \lambda_2 +\lambda_1 \lambda_3+ \lambda_2\lambda_3\,, \label{def3}
\end{eqnarray}
which are combinations appearing in (\ref{poly3}).
This rewriting allows us to rewrite the cubic equation in terms of the universal invariants $i_1,\ldots, i_5$, without using the $i_2^{s,t,u}$ invariants, which may not exist for an arbitrary model.
The universal cubic equation now reads
\begin{equation}
\lambda^3-i_2 \lambda^2 +\frac{i_5}{4}\lambda-\frac{i_3^2+i_4^2}{4i_1}=0\,
\label{cubicequation}
\end{equation}
and its roots are given by
\begin{eqnarray}
\lambda_1 &=&\frac{i_2}{3}+\frac{u}{3w}+\frac{w}{3},\\[2mm]
\lambda_2 &=&\frac{i_2}{3}-\frac{(1-i \sqrt{3})w}{6}-\frac{(1+i\sqrt{3})u}{6w},\\[2mm]
\lambda_3 &=&\frac{i_2}{3}-\frac{(1+i \sqrt{3})w}{6}-\frac{(1-i\sqrt{3})u}{6w}\,,
\end{eqnarray}
where
\begin{eqnarray}
u&=&i_2^2-\frac{3i_5}{4},\\[2mm]
v&=&2i_2^3+\frac{27(i_3^2+i_4^2)}{4i_1}-\frac{9i_2i_5}{4}\, ,\\[2mm]
z&=&\frac{9i_2(i_3^2+i_4^2)i_5}{8i_1}+\frac{i_2^2i_5^2}{16}-\frac{i_2^3(i_3^2+i_4^2)}{i_1}-\frac{27(i_3^2+i_4^2)^2}{16i_1^2}-\frac{i_5^3}{16}\, ,\\[2mm]
w&=&\left(\frac{v+3i\sqrt{3z}}{2}\right)^{1/3}\, .
\end{eqnarray}
As it turns out (by direct evaluation in a specific model, i.e.~the quadratic series) that $u\geq 0$, $v\geq 0$, and since only  $z\geq 0$ is compatible with real roots, one has the relation $ww^*=u$ and thus the above formulae simplify to
\begin{eqnarray}
\lambda_1 &=&\frac{1}{3}\left(i_2+ \, {\rm Re}\, w\right), \label{sollambda1}\\[2mm]
\lambda_2 &=&\frac{1}{3}\left(i_2-{\rm Re}\, w-\sqrt{3}\,{\rm Im}\, w\right),\\[2mm]
\lambda_3 &=&\frac{1}{3}\left(i_2-{\rm Re}\, w+\sqrt{3}\,{\rm Im}\, w\right) . \label{sollambda3}
\end{eqnarray}
Taking into account that $i_1>0$, $i_2>0$, $i_5>0$, the roots are not only real, but also positive.

The outcome of this analysis is an expression of the cubic roots in any arbitrary model and hence this allows us to write down the universal ``fake'' superpotential:
we take the function $W$ of the $stu$ model and replace the model specific invariants $i_2^s, i_2^t, i_2^u$ by $\lambda_1$, $\lambda_2$ and $\lambda_3$, according to their expression computed in (\ref{sollambda1})--(\ref{sollambda3}):
\begin{equation}
	W(i_1,i_2^s,i_2^t,i_2^u,I_4) \longrightarrow W(i_1, \lambda_1, \lambda_2, \lambda_3,I_4).
\end{equation}

We would like now to show that (\ref{cubicequation}), together with the definitions (\ref{def1})--(\ref{def3}), embraces all possible cases, i.e.~all symmetric geometries and all attractor solutions.
As a first example consider the quadratic series for which $C_{ijk} = 0$, then $i_3 = i_4 = i_5 = 0$, so that $\lambda = i_2$.
The superpotential is obviously $W = \sqrt{i_2}$ and the non-BPS attractor is given by $i_1 = 0$.
For cubic geometries, the non-BPS attractor with $Z \neq 0$ occurs at
\begin{equation}
	i_2 = 3 i_1, \quad i_3 = 0, \quad i_4 = -2 i_1^2, \quad i_5 = 12 i_1^2.
\end{equation}
At these points $\lambda_1 = \lambda_2 = \lambda_3 = \frac13 i_2 = i_1$.
On the other hand, for the $Z = 0$ attractor point we have
\begin{equation}
	i_1 = 0, \quad i_3=i_4=i_5 = 0, \quad \lambda_1 = i_2, \quad \lambda_2 = \lambda_3 = 0.
\end{equation}
The superpotential in this case is $W = \sqrt{\lambda_1}$, where $\lambda_1$ is the highest of these eigenvalues.
There is one exception to this case and it is given by the cubic series based on the special geometries 
\begin{equation}
	\frac{{\rm SU}(1,1)}{{\rm U}(1)} \times \frac{{\rm SO}(2,n)}{{\rm SO}(2) \times {\rm SO}(n)}.
\end{equation} 
When $n = 2$, we have the $stu$ model and then there is a complete symmetry between $\lambda_1$, $\lambda_2$ and $\lambda_3$, so that there are 3 different branches of $Z = 0$ black holes depending on which $\lambda_i \neq 0$ at the attractor point. 
For $n>2$ there are still two branches \cite{orbits2} and it can be seen by inspecting the cubic equation (\ref{cubicequation}), which in this case reduces to
\begin{equation}
	\lambda^3 - \lambda^2 (i_2^s + Z_a \bar Z^a)+ \lambda \left(i_2^s\, Z_a \bar Z^a + \frac14 \left|{\rm e}^{K_0}\delta_{\bar a \bar b}Z^{\bar a} Z^{\bar b}\right|^2\right) - \frac{i_2^s}{4}\left|{\rm e}^{K_0}\delta_{\bar a \bar b}Z^{\bar a} Z^{\bar b}\right|^2 = 0,
\end{equation}
where $s$ is the modulus of the first factor and $a,b$ are indices parameterizing the directions associated to the other moduli, using the symplectic sections
\begin{equation}
	X^0 = \frac12(1+ y^2), \quad X^1 = \frac{i}{2}(1-y^2), \quad X^a = y^a, \qquad F_\Lambda = s \, \eta_{\Lambda \Sigma} X^\Sigma \quad (\eta = \{++-...-\}).
\end{equation}
Note that these are not special coordinates ($s$ does not appear in the parameterization of $X^\Lambda$), and that in this case $C_{sab} = -{\rm e}^K \delta_{ab}$ \cite{N2D4}, where the K\"ahler potential $K = K_s + K_0$, due to the factorized form of the scalar manifold.
The cubic equation factorizes as
\begin{equation}
	\left(\lambda- i_2^s\right)\left(\lambda^2 - a \lambda + b\right) = 0,
\end{equation}
where $a \equiv Z_a \bar Z^a $ and $b \equiv \frac14 \left|{\rm e}^{K_0}\delta_{\bar a \bar b}Z^{\bar a} Z^{\bar b}\right|^2$.
Therefore the roots are $\lambda_1 = i_2^s$ and 
\begin{equation}
	\lambda_{2,3} = \frac12 \, Z_a \bar Z^a  \pm \frac12 \sqrt{\left(Z_a \bar Z^a\right)^2 - \left|{\rm e}^{K_0}\delta_{\bar a \bar b}Z^{\bar a} Z^{\bar b}\right|^2}.
	\label{rf2}
\end{equation}
The two $Z=0$ branches of the black hole solutions in this context are therefore 
\begin{equation}
	W = \sqrt{i_2^s}
	\label{rf1}
\end{equation}
and
\begin{equation}
	W = \sqrt{\lambda_2}.
\end{equation}
On the first branch the attractor point follows from $Z_a = 0$ and hence $\lambda_2 = \lambda_3 = 0$, while on the second branch $i_2^s = 0$ and $\delta_{a b}\bar Z^{a} \bar Z^{b} = 0$  and hence $\lambda_3 = 0$.

We point out that (\ref{rf1}) was already obtained in our previous paper \cite{Ceresole:2009iy} and (\ref{rf2}) was also obtained in \cite{Bossard:2009we}.

\subsection{Connection with $N=8$}

In the previous section we have discussed how $W$, when expressed in terms of U-duality invariants, may enjoy a universal form that can be extrapolated from the $stu$ model to all symmetric theories in $N=2$.
We now argue that we can further extend the reasoning to $N=8$ theories.
More precisely, we can view the $stu$ model as a special truncation of $N=8$ supergravity, where the central charge $Z_{AB}$ is diagonal
\begin{equation}
\left(
        \begin{array}{cccc}
          Z \epsilon& & & \\
         &  -i(g^{s\bar s})^{1/2}{\bar D}_{\bar s}{\bar Z}
\epsilon & & \\
         & &- i(g^{t\bar t})^{1/2}{\bar D}_{\bar t}{\bar Z}
\epsilon & \\
        &&&  -i(g^{u\bar u})^{1/2}{\bar D}_{\bar u}{\bar Z}
\epsilon  \\
        \end{array}
      \right)\,.
\end{equation}
Here $\epsilon$ is the 2-dimensional antisymmetric tensor.

This identification allows us to relate the $N=2$ attractor equations of the $stu$ model to the $N=8$ ones and eventually the $N=2$ invariants with the $N=8$ eigenvalues of the central charge.
Because of the factorized structure of the $stu$ model, the attractor equations \cite{Kallosh:2006bt}
\begin{equation}
2{\bar Z}D_iZ+i g^{j\bar l}C_{ijk}g^{k\bar k}{\bar D}_{\bar k}{\bar Z} {\bar
D}_{\bar j} {\bar Z} =0
\end{equation}
can be rewritten as a set of equations
\begin{eqnarray}
2{\bar Z}D_t Z &=&-2 i C_{tsu}g^{s \bar s}g^{u\bar u}{\bar D}_{\bar s}
{\bar Z}{\bar D}_{\bar u} {\bar Z} \nonumber\\[2mm]
&=&-2 i (g_{t\bar t})^{1/2}(g^{s\bar s})^{1/2}(g^{u\bar u})^{1/2}{\bar
D}_{\bar s}{\bar Z}{\bar D}_{\bar u} {\bar Z}\\[2mm]
&\rightarrow& {\bar Z}(g^{t\bar t})^{1/2}D_t Z=-i(g^{s\bar s})^{1/2}
{\bar D}_{\bar s}{\bar Z}(g^{u\bar u})^{1/2}{\bar D}_{\bar u} {\bar Z}\\[2mm]
&\rightarrow& {Z}(g^{t\bar t})^{1/2}{\bar D}_{\bar t} {\bar Z}=-i(g^{s \bar s})^{1/2}{ D}_{ s}{ Z}(g^{u\bar u})^{1/2}{ D}_{ u} {Z}\,,
\end{eqnarray}
which coincide with the $N=8$ algebraic attractor equations
\begin{equation}
z_1 z_2+z_3^\ast z_4^\ast=0, \quad \qquad 1\to 2\to 3\to 4,
\end{equation}
for the four complex eigenvalues $z_i$ of $Z_{AB}$ \cite{Ferrara:2006em}.
The relation between the two quantities is
\begin{equation}
Z=z_1,\quad \quad {\bar Z}^s=i\, z_2\quad , \quad{\bar Z}^t=i\, z_3\quad , \quad{\bar Z}^u=i \,z_4\,.
\end{equation}
We can therefore elaborate on the relation between the sets of eigenvalues of the $N=8$ central charge and the $N=2$ invariants by bringing the four complex eigenvalues to the normal form $z_i=\rho_i{\rm e}^{i\phi/4}$ by means of an SU(8) transformation and comparing the $N=8$ quartic invariant with the $N=2$ quartic invariant.
The $N=8$ quartic invariant reads \cite{Kallosh:1996uy,Ferrara:1997ci}
\begin{eqnarray}
I_4&=&\rho_1^4+\rho_2^4+\rho_3^4+\rho_4^4-2(\rho_1^2\rho_2^2+
\rho_1^2\rho_3^2+\rho_1^2\rho_4^2+\nonumber\\[2mm]
&&\rho_2^2\rho_3^2+\rho_2^2\rho_4^2+ \rho_3^2\rho_4^2)+8\rho_1\rho_2\rho_3\rho_4 \cos\phi.
\end{eqnarray}
At the non-BPS attractor point $\rho_1=\rho_2=\rho_3=\rho_4 = \rho$ and $\phi = \pi$, so that
\begin{equation}
	I_4 = -16 \rho^4.
\end{equation}
This can be compared with the corresponding expression for the $N=2$ quartic invariant, which reads
\begin{eqnarray}
I_4 &=& (Z{\bar Z}-Z_i{\bar Z}^i)^2+\frac23 i[Z N_3({\bar Z})-{\bar
Z}N_3(Z)]\nonumber\\[2mm]
&-&C_{ijk}C_{{\bar \imath}{\bar \jmath}{\bar k}}g^{i\bar \imath}{\bar Z}^j{\bar Z}^k Z^{\bar \jmath} Z^{\bar k}.
\end{eqnarray}
At this stage we have 4 real eigenvalues and a phase in the $N=8$ case, versus 5 invariants in $N=2$ models.
These can also be assembled in 4 invariants, one per $N=8$ eigenvalue, and a phase $\theta$, introduced in analogy with $N=8$ by defining
\begin{equation}
e^{2i\theta}\equiv-\frac{Z N_3({\bar Z})}{{\bar Z}{\bar N}_3(Z)}.
\end{equation}
We expect that the general ``fake'' superpotential for the full $N=8$ theory can be obtained by taking the superpotential of the $stu$ model $W_{stu}(i_1, i_2^s, i_2^t, i_2^u, i_4, I_4)$ and replacing the first 4 invariants by the generic square eigenvalues of $Z_{AB}$, $(\rho_1^2,\rho_2^2,\rho_3^2, \rho_4^2)$, and further replacing the quartic invariant $I_4$ of $N=2$ with $I_4$ of the  $E_{7(7)}$ scalar manifold, and $i_4$ by 2 Re Pf $Z$.

Adapting the previous formulae to the $stu$ model we find
\begin{eqnarray}
I_4 &=&(i_1-i_2^s-i_2^t-i_2^u)^2+4i_4-4(i_2^s i_2^t+i_2^s i_2^u+i_2^t i_2^u)=\\
&&i_1^2+(i_2^s)^2+(i_2^t)^2+(i_2^u)^2+4i_4-2(i_1 i_2^s+i_1 i_2^t+i_1 i_2^u+i_2^s i_2^u+i_2^s i_2^t+i_2^t i_2^u)
\end{eqnarray}
where $4i_4=8$ Re Pf $Z$ and
\begin{equation}
i_4=i (Z{\bar Z}^s{\bar Z}^t {\bar Z}^u-{\bar Z}Z^{\bar s} Z^{\bar t} Z^{\bar u}),
\end{equation}
which, in terms of the $N=8$ quantities can be rewritten as
\begin{equation}
4i_4=4(z_1 z_2 z_3 z_4+{\bar z}_1{\bar z}_2 {\bar z}_3 {\bar z}_4)=8
{\rm Re} \, {\rm Pf}\, Z\, .
\end{equation}
Note that $I_4$ not only has the triality typical of the {\it stu} model \cite{triality,Behrndt:1996hu}, but also a quadrality that involves also $i_1$, as is evident
from $V_{BH}$ and from the particular solution at $i_3=0$:
\begin{equation}
W=\frac12(\sqrt{i_1}+\sqrt{i_2^s}+\sqrt{i_2^t}+\sqrt{i_2^u})= \frac12\left(\rho_1+\rho_2+\rho_3+\rho_4\right)\, .
\end{equation}

The overall phase of $ZN_3({\bar Z})$ makes it purely imaginary, so that $\theta=\pi/2$ ($I_4<0$).
At this point Pf $Z$ is real and negative and so its phase is $\pi$.
Then we have $\phi=\theta+\frac{\pi}{2}$.
The precise relation comes from the fact that at the attractor point 
\begin{equation}
I_4^{N=8} \supset 8 {\rm Re}\ {\rm Pf}\ Z=8 \rho_1 \rho_2 \rho_3 \rho_4 \cos \phi=(-8 \rho^4),
\end{equation}
while the $N=2$ quartic invariant contains
\begin{equation}
I_4^{N=2} \supset 4 i_4 = \frac{2}{3}i\left[Z N_3 ({\bar Z})-{\bar
Z}N_3(Z)\right]= -8 i_1^2,
\end{equation}
which means that
\begin{equation}
8 {\rm Re}\ {\rm Pf}\ Z=4 i_4\Rightarrow {\rm Re}\ {\rm Pf}\ Z=\frac12 i_4<0.
\end{equation}
Summarizing, the black hole solutions of the $N=8$ theory should follow by replacing
\begin{eqnarray}
&&i_4 \rightarrow 2 {\rm Re}\ {\rm Pf}\ Z_{AB}\nonumber\\
&&i_1,  i_2^s, i_2^t, i_2^u \rightarrow {\rm eigenvalues}\ {\rm of}\ Z_{AB}{\bar Z}^{BC},
\end{eqnarray}
where the eigenvalues are given by the four roots of the equation
\begin{equation}
\lambda^4+a\,\lambda^3+b\,\lambda^2+c\,\lambda+d=0,
\end{equation}
where
\begin{eqnarray}
a&=&-\frac12 Tr Z Z^\dagger,\nonumber\\[2mm]
b&=&\frac14\left[\frac12 (Tr Z Z^\dagger)^2-Tr(Z Z^\dagger)^2\right],\nonumber\\[3mm]
c&=&-\frac16\left[\frac18(Tr ZZ^\dagger)^3+Tr(ZZ^\dagger)^3-\frac34Tr\ ZZ^
\dagger Tr(ZZ^\dagger)^2\right],\nonumber\\[3mm]
d&=&\frac14\left\{\frac{1}{96}(TrZZ^\dagger)^4+\frac18[Tr(ZZ^\dagger)^2]^2+
\frac13Tr(ZZ^\dagger)^3TrZZ^\dagger\right.\nonumber\\[2mm]
&-&\left.\frac12Tr(ZZ^\dagger)^4-\frac18(TrZZ^\dagger)^2Tr(ZZ^\dagger)^2\right\},
\end{eqnarray}
as given in \cite{D'Auria:1999fa}.
This gives $W(\rho_1,\rho_2,\rho_3,\rho_4,{\rm Re}\ {\rm Pf}\ Z)$.

Summing up, the claim is the following: once we know the expression for $W$ in the {\it stu} model, we have
\begin{equation}
W(\rho_1,\rho_2,\rho_3,\rho_4,\phi)
\end{equation}
where $\rho_1=i_1$, $\rho_2=i_2^s$, $\rho_3=i_2^t$, $\rho_4=i_2^u$, and  $ \phi$ is the phase of the ${\bar Z} N_3(Z)$ object related to the phase of
the $N=8$ given above. Since $\rho_i,\phi$ are generic, any other model should be obtained by replacing $\rho_i,\phi$ with the objects of the
other model ($\phi$ can also be expressed through $\rho_i$ and $I_4$).

\section{Concluding Remarks} 
\label{sec:concluding_remarks}

This paper provides a general technique to explicitly construct the ``fake'' superpotential $W$ driving the gradient flow equations for non-BPS black holes in $N=2$ supergravities based on symmetric spaces.
The procedure consists of four clear steps, starting with: a) the determination of $W$ for a simple charge configuration by using the charge rotation technique of \cite{Ceresole:2007wx}; b) generalizing the result to an arbitrary charge configuration by a duality rotation; c) rewriting the resulting function in terms of duality invariants, treating the non-invariant combinations as auxiliary fields and finally d) integrating these auxiliary fields out.
We have successfully applied this procedure to the $st^2$ and $stu$ cases, also discussing how the same procedure can be generalized to any arbitrary model in $N=2$ and $N=8$.

An alternative general method to obtain the same set of solutions has been provided in \cite{Bossard:2009we}, where the ``fake'' superpotential functions for non-BPS black holes in the context of the $N=8$ theory have been obtained in an implicit form through the analysis of geodesic equations of the timelike reduced 3-dimensional models.
Although the two procedures should obviously lead to the same results, we emphasize that the use of auxiliary fields allow for an explicit description of the $W$ functions. For this reason, in our approach we are able to provide full analytic solutions to the flows.
It is only after  eliminating all the auxiliary fields ( by the solution of the constraint equations following from the variation of the superpotential with respect to them) that our expression for $W$ and the one of \cite{Bossard:2009we} will agree.
This, however, requires in both  cases the solution of a polynomial equation that generically is of degree 6, and hence remains implicit for us as well.
We recall that the power of the first order formalism and the main motivation behind its construction is the fact that, for a given $W$, one is not only able to obtain the horizon properties of a given non-BPS black hole solution, but also to construct the full solution, from asymptotic infinity to the horizon.
In this respect we think that our approach is better suited for the  construction of new solutions, because it always allows for an explicit construction of $W$ by using an appropriate number of auxiliary fields.

We did not discuss here $N=2$ models that are not based on coset scalar manifolds.
It would be interesting to understand better how to apply our procedure to these cases, where a general classification of the duality invariant quantities has not been performed yet.
Finally, we also expect that our approach could be easily generalized to the 5-dimensional black hole solutions, for which the first order formulation was already used to derive new non-BPS solutions \cite{Cardoso5d}.


\bigskip
\section*{Acknowledgments}

\noindent A.Y. would like to thank A Shcherbakov for discussions and useful comments.
This work is supported in part by the ERC Advanced Grant no. 226455, \textit{``Supersymmetry, Quantum Gravity and Gauge Fields''} (\textit{SUPERFIELDS}).
The work of A.~C.~is partially supported by MIUR-PRIN contract 20075ATT78, the work of G.~D.~has been partially supported by the Fondazione Cariparo Excellence Grant {\em String-derived supergravities with branes and fluxes and their phenomenological implications} and by the European Programme UNILHC (contract PITN-GA-2009-237920), the work of S.~F.~has been supported in part by D.O.E.~grant DE-FG03-91ER40662, Task C and the work of A.~Y.~has been supported in part by CERN-PH-TH where part of the work was done.


\begin{thebibliography}{99}
	
\bibitem{SUSY} 	 S.~Ferrara, R.~Kallosh and A.~Strominger, \textit{``N=2 extremal black holes,''} Phys.\ Rev.\ D {\bf 52} (1995) 5412 [arXiv:hep-th/9508072]; \\
A. Strominger, \textit{Macroscopic entropy of }$\mathcal{N}\mathit{=2}$\textit{\ extremal black holes}, Phys. Lett. \textbf{B383}, 39 (1996), [arXiv:hep-th/9602111];
S. Ferrara and R. Kallosh, \textit{Supersymmetry and attractors}, Phys. Rev. D \textbf{54}, 1514 (1996), [arXiv:hep-th/9602136];
S. Ferrara, R. Kallosh, \textit{Universality of supersymmetric attractors}, Phys. Rev. D\textbf{54}, 1525 (1996), [arXiv:hep-th/9603090];
S. Ferrara, G. W. Gibbons and R. Kallosh, \textit{Black holes and critical points in moduli space}, Nucl. Phys. B \textbf{500}, 75 (1997), [arXiv:hep-th/9702103].
	
\bibitem{Ceresole:2007wx}  A.~Ceresole and G.~Dall'Agata, \emph{``Flow Equations for Non-BPS Extremal Black Holes,''}  JHEP {\bf 0703} (2007) 110	  [arXiv:hep-th/0702088].

\bibitem{Andrianopoli:2007gt} L.~Andrianopoli, R.~D'Auria, E.~Orazi and M.~Trigiante,
\emph{``First Order Description of Black Holes in Moduli Space,''}	  JHEP {\bf 0711} (2007) 032	 [arXiv:0706.0712 [hep-th]];

\bibitem{Cardoso5d} G.~Lopes Cardoso, A.~Ceresole, G.~Dall'Agata, J.~M.~Oberreuter and J.~Perz,  \emph{``First-order flow equations for extremal black holes in very special geometry,''} JHEP {\bf 0710} (2007) 063 [arXiv:0706.3373 [hep-th]].

\bibitem{varinonBPS} S.~Ferrara, A.~Gnecchi and A.~Marrani, \emph{``d=4 Attractors, Effective Horizon Radius and Fake Supergravity,''}   Phys.\ Rev.\  D {\bf 78} (2008) 065003 [arXiv:0806.3196 [hep-th]];\\
J.~Perz, P.~Smyth, T.~Van Riet and B.~Vercnocke,  \emph{``First-order flow equations for extremal and non-extremal black holes,''}  JHEP {\bf 0903} (2009) 150
 [arXiv:0810.1528 [hep-th]];\\
K.~Goldstein and S.~Katmadas, \emph{``Almost BPS black holes,''}   JHEP {\bf 0905} (2009) 058 [arXiv:0812.4183 [hep-th]];\\
I.~Bena, G.~Dall'Agata, S.~Giusto, C.~Ruef and N.~P.~Warner, \emph{``Non-BPS Black Rings and Black Holes in Taub-NUT,''}  JHEP {\bf 0906} (2009) 015  [arXiv:0902.4526 [hep-th]];\\
P.~Galli and J.~Perz, \emph{``Non-supersymmetric extremal multicenter black holes with superpotentials,''} arXiv:0909.5185 [hep-th].

\bibitem{Bellucci:2008sv}  S.~Bellucci, S.~Ferrara, A.~Marrani and
A.~Yeranyan, \emph{``stu Black Holes Unveiled,''}  arXiv:0807.3503 [hep-th].

\bibitem{Andrianopoli:2009je}  L.~Andrianopoli, R.~D'Auria, E.~Orazi and M.~Trigiante, \emph{``First Order Description of D=4 static Black Holes and the Hamilton-Jacobi equation,''}  arXiv:0905.3938 [hep-th].

\bibitem{Ceresole:2009iy}
A.~Ceresole, G.~Dall'Agata, S.~Ferrara and A.~Yeranyan, \emph{``First order flows for N=2 extremal black holes and duality invariants,''}
arXiv:0908.1110 [hep-th].

\bibitem{Bossard:2009we} G.~Bossard, Y.~Michel and B.~Pioline, \emph{``Extremal black holes, nilpotent orbits and the true fake superpotential,''}  arXiv:0908.1742 [hep-th].

\bibitem{orbits} L.~Andrianopoli, R.~D'Auria and S.~Ferrara, \emph{``U-duality and central charges in various dimensions revisited,''} Int.\ J.\ Mod.\ Phys.\  A {\bf 13}, 431 (1998)
[arXiv:hep-th/9612105];\\
L.~Andrianopoli, R.~D'Auria and S.~Ferrara,\emph{``Five dimensional U-duality, black-hole entropy and topological invariants,''}  Phys.\ Lett.\  B {\bf 411}, 39 (1997)  [arXiv:hep-th/9705024].

\bibitem{orbits2} S.~Bellucci, S.~Ferrara, M.~Gunaydin and A.~Marrani, \emph{``Charge orbits of symmetric special geometries and attractors,''} Int.\ J.\ Mod.\ Phys.\  A {\bf 21} (2006) 5043 [arXiv:hep-th/0606209].

\bibitem{Cremmer:1984hc}  E.~Cremmer and A.~Van Proeyen,  \emph{``Classification Of Kahler Manifolds In N=2 Vector Multiplet Supergravity Couplings,''}  Class.\ Quant.\ Grav.\  {\bf 2}, 445 (1985).

\bibitem{triality} M.~J.~Duff, J.~T.~Liu and J.~Rahmfeld, \emph{``Four-Dimensional String-String-String Triality,'' }Nucl.\ Phys.\ B {\bf 459}, 125 (1996) [arXiv:hep-th/9508094].


\bibitem{Behrndt:1996hu} K.~Behrndt, R.~Kallosh, J.~Rahmfeld, M.~Shmakova and W.~K.~Wong, \emph{``STU black holes and string triality,''} Phys.\ Rev.\ D {\bf 54} (1996) 6293 [arXiv:hep-th/9608059]. 

\bibitem{Ferrara:2007tu} S.~Ferrara and A.~Marrani, \emph{``On the Moduli Space of non-BPS Attractors for N=2 Symmetric Manifolds,''}  Phys.\ Lett.\  B {\bf 652} (2007) 111 [arXiv:0706.1667 [hep-th]].

\bibitem{Ferrara:1997ci} S.~Ferrara and J.~M.~Maldacena, \emph{``Branes, central charges and U-duality invariant BPS conditions,''} Class.\ Quant.\ Grav.\  {\bf 15}, 749 (1998) [arXiv:hep-th/9706097].

\bibitem{D'Auria:1999fa} R.~D'Auria, S.~Ferrara and M.~A.~Lledo, ``On central charges and Hamiltonians for 0-brane dynamics,'' Phys.\ Rev.\  D {\bf 60} (1999) 084007
[arXiv:hep-th/9903089].

\bibitem{Ceresole:2007rq} A.~Ceresole, S.~Ferrara and A.~Marrani, \emph{``4d/5d Correspondence for the Black Hole Potential and its Critical Points,''}  Class.\ Quant.\ Grav.\  {\bf 24}, 5651 (2007) [arXiv:0707.0964 [hep-th]].

\bibitem{Sen:1994eb} A.~Sen, \emph{``Black Hole Solutions In Heterotic String Theory On A Torus,''} Nucl.\ Phys.\  B {\bf 440} (1995) 421  [arXiv:hep-th/9411187].

\bibitem{Cvetic:1996zq}  M.~Cvetic and C.~M.~Hull, \emph{``Black holes and U-duality,''}  Nucl.\ Phys.\  B {\bf 480} (1996) 296   [arXiv:hep-th/9606193].

\bibitem{Cvetic:1995uj}  M.~Cvetic and D.~Youm, \emph{``Dyonic BPS saturated black holes of heterotic string on a six torus,''}   Phys.\ Rev.\  D {\bf 53} (1996) 584  [arXiv:hep-th/9507090].

\bibitem{Gimon:2007mh}  E.~G.~Gimon, F.~Larsen and J.~Simon, \emph{``Black Holes in Supergravity: the non-BPS Branch,''}  JHEP {\bf 0801} (2008) 040  [arXiv:0710.4967 [hep-th]].

\bibitem{Cerchiai:2009pi} B.~L.~Cerchiai, S.~Ferrara, A.~Marrani and B.~Zumino, \emph{``Duality, Entropy and ADM Mass in Supergravity,''} Phys.\ Rev.\  D {\bf 79} (2009) 125010 [arXiv:0902.3973 [hep-th]].

\bibitem{Gunaydin:1983bi}  M.~Gunaydin, G.~Sierra and P.~K.~Townsend,
\emph{``The Geometry Of N=2 Maxwell-Einstein Supergravity And Jordan Algebras,''}  Nucl.\ Phys.\ B {\bf 242} (1984) 244.

\bibitem{Strominger:1990pd}  A.~Strominger,\emph{``Special Geometry,''}  Commun.\ Math.\ Phys.\ {\bf 133} (1990) 163.

\bibitem{Kallosh:1996uy} R.~Kallosh and B.~Kol, \emph{``E(7) Symmetric Area of the Black Hole Horizon,''} Phys.\ Rev.\  D {\bf 53}, 5344 (1996)
[arXiv:hep-th/9602014].

\bibitem{Ferrara:2006em}  S.~Ferrara and R.~Kallosh, \emph{``On N = 8 attractors,''}  Phys.\ Rev.\  D {\bf 73} (2006) 125005  [arXiv:hep-th/0603247].

\bibitem{Kallosh:2006bt} R.~Kallosh, N.~Sivanandam and M.~Soroush,
\emph{``The non-BPS black hole attractor equation,''} JHEP {\bf 0603}, 060 (2006) [arXiv:hep-th/0602005]. 

\bibitem{N2D4}  L.~Andrianopoli, M.~Bertolini, A.~Ceresole, R.~D'Auria, S.~Ferrara, P.~Fre and T.~Magri,  \emph{``N = 2 supergravity and N = 2 super Yang-Mills theory on general scalar
    manifolds: Symplectic covariance, gaugings and the momentum map,''}J.\ Geom.\ Phys.\  {\bf 23} (1997) 111  [arXiv:hep-th/9605032].

\end{thebibliography}
\end{document}